# Surface plasmon induced quantum interference at meta-material interface

Ali A. Kamli [(1)], Jabir Hakami [(1)], and M. Suhail Zubairy [(2)]

[(1)] Department of Physics, Jazan University, Jazan Box 114, Saudi Arabia,

[(2)] Institute for Quantum Science and Engineering (IQSE), Department of Physics & Astronomy, Texas A&M University, College Station, Texas 77845 USA

Abstract

In this work we investigate quantum interference in a four-level atom coupled to a negative index meta-material (NIMM) anisotropic plasmonic environment that supports both TE and TM polarized surface plasmons (SP). The analysis confirms the creation of the anisotropic environment and two dipoles can interfere with each other even if they are orthogonal by sharing such SP modes. The NIMM/plasmonic environment provides more options to control SP interaction with emitters and their spontaneous emission decays and spectrum. The spectrum depends critically on structure parameters, mode frequency, frequency dependent electric permittivity and magnetic permeability, and the location of the atom etc. We observe orders of magnitudes enhancement in the plasmonic-modified decays and spectrum compared to free space case.

1. **Introduction**

Quantum interference (QI) and its control in atomic systems proved useful tools for deeper understanding of quantum physics and for technological applications. Interesting and novel QI features in the spectrum of such systems were reported such as lasing without inversion, electromagnetically induced transparency, dark states, spectral line broadening and narrowing, line elimination just to name a few [1-11]. In multi-level atomic systems, QI occurs when there are two or more competing channels for spontaneous decays. Early research [7-11] on quantum interference was conducted in isotropic free space with the condition that the dipole matrix elements of the competing channels must be parallel or anti- parallel, which is hard to achieve in experiments as dipoles are oriented arbitrary and radiation is emitted in random directions. To overcome this problem, Agarwal [12] suggested working with anisotropic environment and demonstrated existence of QI even when the dipole moments are orthogonal. Since then much work was carried out to explore QI effects in different anisotropic media [13-25]. A control of quantum interference can be achieved through phase or/and amplitude of driving fields in multilevel atoms, or by coupling the atomic system to tailored environments, or cavity systems that can be controlled at will by modification of spontaneous emission of the competing channels that produce quantum interference. Nano-photonics technology, on the other hand, requires strong atom-field coupling, which can be achieved via different environments such as in a Fabry Perot cavity, the photonic crystals [13], nanostructures and optical fibers [18-20], graphene structures [21], and surface plasmons [22-25].

In this work, we consider a plasmonic system [26-30] that supports surface plasmon modes (SP). These SP modes couple strongly to emitters due to their highly confined fields, thus increasing the atom-field interaction, which in turn enhances spontaneous emission process that causes QI. To be more specific, we are interested to study control of quantum interference in a driven four level system induced by SP modes that arise at the interface between a dielectric and a negative index meta-material (NIMM) [31-37]. NIMM are artificially fabricated materials that support both transverse electric (TE) and transverse magnetic (TM) polarized SP modes with low loss and energy confined to the interface.

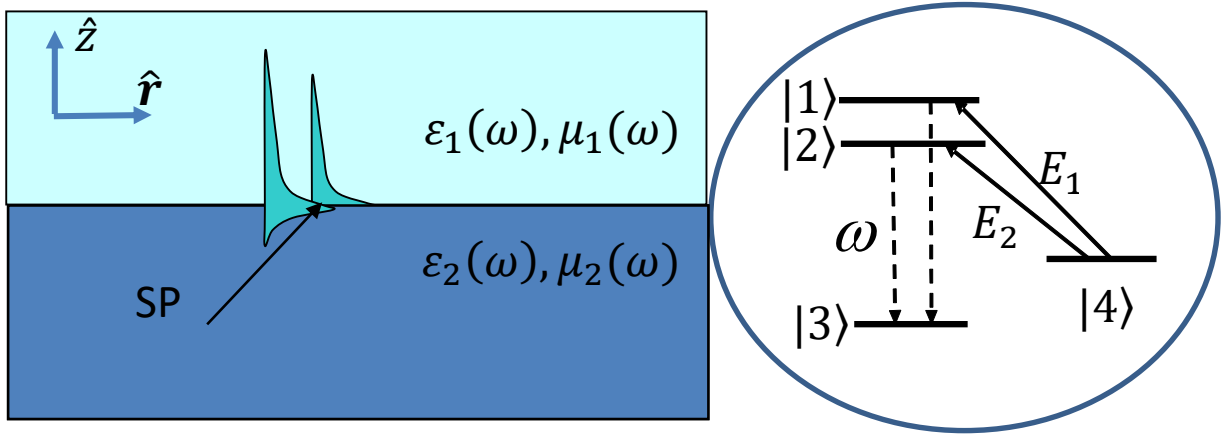

Fig.1: Left: Plasmonic system supporting SP modes consisting of upper half space (z > 0 ) of permittivity $\varepsilon_1(\omega)$ and permeability $\mu_1(\omega)$ and lower half space ( z < 0) of permittivity $\varepsilon_2(\omega)$ and permeability $\mu_2(\omega)$. The two media are joined at interface z=0. SP modes propagate at interface along the in-plane wave vector $\boldsymbol{r}$, and decay along z-direction. Right: The four-level system (levels |j>, j=1-4) interacting with SP modes of frequency $\omega$ and with fields $E_1$ and $E_2$, is placed above interface at position z.

The interest in these artificially fabricated materials has increased by the fabrication of NIMMs at optical and desirable frequencies, the emergence of quantum meta-materials, and other applications that include cloaking, super resolution images and the possibilities of implementing in optical quantum processing [38–41].

This paper is organized as follows. In section 2, we present the plasmonic system at the boundary of a dielectric and negative index meta-material and introduce the necessary formalism. In section 3, we quantize these SP fields, and section 4 considers their interactions with a four-level atomic system where we solve the equations of motion and explore QI effects. In section 5, we present results and conclude in section 6 with summary and discussion.

## 2. Plasmonic Environment

In figure 1, we show the system that generates the physical plasmonic environment. It consists of two half spaces joined at the interface z=0 in the x-y plane. The upper half space (z>0) is taken to be air or any dielectric material characterized by constant dielectric function or permittivity $\varepsilon_1$ and constant magnetic permeability $\mu_1$. The lower half space (z<0) medium is a negative index metamaterial (NIMM) characterized by frequency dependent complex dielectric function or permittivity $\varepsilon_2(\omega)$ and complex magnetic permeability $\mu_2(\omega)$. For certain frequency range, the real parts of $\varepsilon_2(\omega)$ and $\mu_2(\omega)$ are negative and both transverse electric (TE) and transverse magnetic (TM) polarized plasmonic modes can exist simultaneously. The two-dimensional SP electric fields are confined to the interface plane and propagate in the x-y plane with in-plane complex wave vector $\boldsymbol{K}_\| = (K_x, K_y)$, with field amplitudes decay away in both sides with distance from the interface at $z = 0$. So the SP electric field $\boldsymbol{E}$ of a transverse mode of frequency $\omega$ satisfying the wave equation $\nabla^2 \boldsymbol{E}_m + \omega^2 \varepsilon_o \mu_o \varepsilon_m(\omega) \mu_m(\omega) \boldsymbol{E}_m = 0$ (m=1, 2 for the two media), is of the form $\boldsymbol{E}_1 = \boldsymbol{A}_1 e^{i(\boldsymbol{K}_\| \cdot \boldsymbol{r}_\| - \omega t)} e^{-k_1 z}$ in the upper half space (z > 0), and $\boldsymbol{E}_2 = \boldsymbol{A}_2 e^{i(\boldsymbol{K}_\| \cdot \boldsymbol{r}_\| - \omega t)} e^{k_2 z}$ in the lower space (z < 0), and the in-plane vector $\boldsymbol{r}_\| = (x, y)$. Here the constants $A_1$ and $A_2$ can be determined from boundary conditions. Where $\varepsilon_0$ is the vacuum dielectric constant (or permittivity) and $\mu_0$ is the vacuum permeability, $c = 1/\sqrt{\varepsilon_0 \mu_0}$ is speed of light in vacuum. The wave numbers $k_m = \sqrt{K_\|^2 - (\omega^2/c^2)\varepsilon_m(\omega)\mu_m(\omega)}$ are the wave vector components along z-direction normal to the interface characterized by positive real parts $\text{Re}[k_m] > 0$ so that the SP field amplitudes decay away from interface. These SP modes are thus bound to interface and propagate at wave vector $K_\|$ parallel to interface. Applications of appropriate electromagnetic boundary conditions at interface z=0, leads to the following conditions [27, 28]

$$k_1 \mu_2(\omega) + k_2 \mu_1(\omega) = 0 \quad , \quad K_\| = k_\| + i\kappa = \frac{\omega}{c}\sqrt{\mu_1 \mu_2 \frac{\varepsilon_1 \mu_2 - \varepsilon_2 \mu_1}{\mu_2^2 - \mu_1^2}}, \quad (1)$$

for TE polarized SP modes, and

$$k_1 \varepsilon_2(\omega) + k_2 \varepsilon_1(\omega) = 0 \quad , \quad K_\| = k_\| + i\kappa = \frac{\omega}{c}\sqrt{\varepsilon_1 \varepsilon_2 \frac{\mu_1 \varepsilon_2 - \mu_2 \varepsilon_1}{\varepsilon_2^2 - \varepsilon_1^2}} \quad (2)$$

for TM polarized SP modes. The real part $k_\parallel$ of the complex wave vector $K_\parallel$ gives the dispersion relations, while the imaginary part $\kappa$ gives SP loss that determines the SP propagation distance along the interface. The positive real parts of the wave numbers $k_m$, normal to interface give the skin or penetration depth of the fields into both media, which we take as our definition of field confinement and denote as $\zeta_m = 1/Re[k_m]$. Since real $k_{1,2}$ are positive, Eq. (1) is fulfilled when the magnetic permeability of one of the two media has negative real part and similarly for Eq. (2). Thus in a NIMM, where real parts of both electric permittivity and magnetic permeability are negative, equations 1 and 2 can be satisfied simultaneously and both transverse magnetic and electric surface plasmons can exist at the same time. The first medium is described by the pair ($\mu_1 = 1$ and $\varepsilon_1 = 1$), while NIMM is modeled in the Drude model by the frequency dependent electric permittivity $\varepsilon_2(\omega)$, and magnetic permeability $\mu_2(\omega)$ [32-35] as;

$$\varepsilon_2(\omega) = 1 - \frac{\omega_e^2}{\omega(\omega + i\gamma_e)} \quad , \quad \mu_2(\omega) = 1 - \frac{\omega_h^2}{\omega(\omega + i\gamma_h)} \quad , \qquad (3)$$

where $\omega_e$ the plasma frequency usually in the ultraviolet region is, $\gamma_e$ is the electric damping rate due to material losses, $\omega_h$ is the magnetic plasma frequency, and $\gamma_h$ is the magnetic damping rate. The dispersions and losses are shown in Fig. 2, for the set of parameters; $\omega_e = 1.37 \times 10^{16} s^{-1}$, $\gamma_e = 2.73 \times 10^{15} s^{-1}$ and since the medium response to the magnetic component of the field is weaker than the electric component, we assume $\omega_h = 0.6\omega_e$ and $\gamma_h = \gamma_e/1000$ [30]. The mode frequency $\omega$ is scaled to plasma frequency $\omega_e$, and wavenumbers $k_\parallel$ and $k_m$ are scaled to $k_e = \omega_e/c = 4.6 \times 10^7 m^{-1}$. Both TM and TE dispersions show foldings at certain frequencies and this will influence the dynamics of SP modes, in particular the behavior of the SP confinement and group velocity, which we shall need in later parts of this work. From Fig. 2b, it is clear that losses are highly reduced for $\omega/\omega_e < 0.42$ and $\omega/\omega_e > 0.42$. The frequency range $0.1 < \omega/\omega_e < 0.7$ which; (1) spans the optical range, (2) includes low loss range (except at frequency $\omega/\omega_e \approx 0.42$ where TE losses are high), and (3) supports both TE and TM modes, will be taken as our working frequency throughout this paper for the set of parameters defined above.

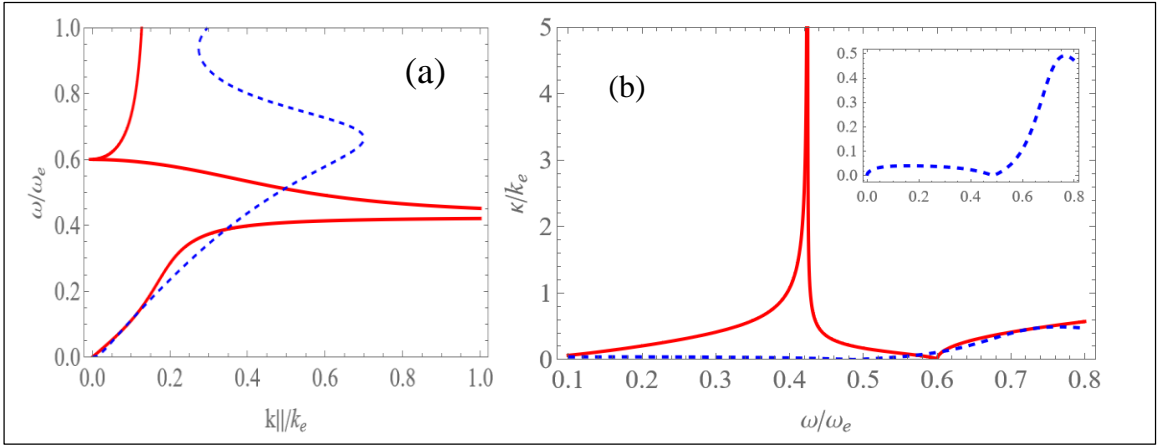

Fig.2. (a) Dispersions and (b) losses given by equations 1 and 2 as functions of mode frequency $\omega/\omega_e$ for TE (solid) and TM (dashed – expanded in the inset) modes scaled to plasma frequency $\omega_e$, and wavenumbers $k_\parallel$ and $\kappa$ are scaled to $k_e = \omega_e/c = 4.6 \times 10^7 m^{-1}$. Foldings in the dispersions occur at $\omega/\omega_e \approx 0.42, 0.60, 0.65$, and losses are reduced for frequencies smaller and bigger than $\omega/\omega_e = 0.42$.

3. **SP quantized fields**

The SP modes constitute the reservoir modes that will couple to emitters, and modify interference effects. In order to quantify these effects, SP modes will be quantized using standard methods. In the low loss range $\mathbf{K}_\parallel \approx \mathbf{k}_\parallel = (k_x, k_y)$, and the quantization procedure (Appendix A) leads to the following expressions for the SP quantized modes;

$$\hat{\mathbf{E}}(\mathbf{r},t) = \frac{A}{4\pi^2} \sum_\alpha \int d^2 \mathbf{k}_\parallel [\mathbf{E}_\alpha(\mathbf{k}_\parallel) a_\alpha(\mathbf{k}_\parallel) e^{i(\mathbf{k}_\parallel \cdot \mathbf{r}_\parallel - \omega t)} + H.C] \qquad (4)$$

where $\alpha$ refers to TE and TM polarized modes. The operators $a_\alpha^+(\mathbf{k}_\parallel)$ and $a_\alpha(\mathbf{k}_\parallel)$ of the plasmonic modes obey the usual equal time commutation relation $[a_\alpha(\mathbf{k}_\parallel), a_{\alpha'}^+(\mathbf{k}_\parallel')] = \delta_{\alpha\alpha'} \delta(\mathbf{k}_\parallel - \mathbf{k}_\parallel')$. The SP field amplitudes $\mathbf{E}_\alpha$ take the form;

$$\boldsymbol{E}_{TE}(\boldsymbol{k}_{\|}) = N_{TE}(\boldsymbol{k}_{\|})(\hat{\boldsymbol{k}}_{\|} \times \hat{\boldsymbol{z}})\left[\theta(z)e^{-k_1 z} + \theta(-z)e^{k_2 z}\right]$$

$$\boldsymbol{E}_{TM}(\boldsymbol{k}_{\|}) = N_{TM}(\boldsymbol{k}_{\|})\left[\theta(z)\,(\hat{\boldsymbol{k}}_{\|} + i\hat{\boldsymbol{z}}\frac{k_{\|}}{k_1})e^{-k_1 z} + \theta(-z)\,(\hat{\boldsymbol{k}}_{\|} - i\hat{\boldsymbol{z}}\frac{k_{\|}}{k_2})e^{k_2 z}\right] \quad (5)$$

where $\hat{\boldsymbol{k}}_{\|} = \boldsymbol{k}_{\|}/k_{\|}$ is a unit vector in the interface plane, $\theta(z)$ is the Heaviside step function and

$$N_\alpha(k_{\|}) = \left|\frac{\hbar\omega(k_{\|})}{\varepsilon_o A L_{z,\alpha}(\omega,\varepsilon,\mu)}\right|^{1/2} \quad \alpha = TE, TM \quad (6)$$

$$L_{z,TE}(\omega,\varepsilon,\mu) = \frac{1}{2}[D_{TE} + \frac{c^2}{\omega^2}S_{TE}]$$

$$D_{TE} = \zeta_1 \operatorname{Re}\left[\frac{\partial}{\partial\omega}(\omega\varepsilon_1)\right] + \zeta_2 \operatorname{Re}\left[\frac{\partial}{\partial\omega}(\omega\varepsilon_2)\right] \quad (7)$$

$$S_{TE} = \zeta_1 \operatorname{Re}\left[\frac{\partial}{\partial\omega}(\omega\mu_1)\right]\frac{|k_1|^2 + |k_{\|}|^2}{|\mu_1|^2} + \zeta_2 \operatorname{Re}\left[\frac{\partial}{\partial\omega}(\omega\mu_2)\right]\frac{|k_2|^2 + |k_{\|}|^2}{|\mu_2|^2}$$

$$L_{z,TM}(\omega,\varepsilon,\mu) = \frac{1}{2}[D_{TM} + \frac{\omega^2}{c^2}S_{TM}]$$

$$D_{TM} = \zeta_1 \operatorname{Re}\left[\frac{\partial}{\partial\omega}(\omega\varepsilon_1)\right]\frac{|k_1|^2 + |k_{\|}|^2}{|k_1|^2} + \zeta_2 \operatorname{Re}\left[\frac{\partial}{\partial\omega}(\omega\varepsilon_2)\right]\frac{|k_2|^2 + |k_{\|}|^2}{|k_2|^2} \quad . \quad (8)$$

$$S_{TM} = \zeta_1 \operatorname{Re}\left[\frac{\partial}{\partial\omega}(\omega\mu_1)\right]\left|\frac{\varepsilon_1}{k_1}\right|^2 + \zeta_2 \operatorname{Re}\left[\frac{\partial}{\partial\omega}(\omega\mu_2)\right]\left|\frac{\varepsilon_2}{k_2}\right|^2$$

The normalization factor $N_\alpha(k_{\|})$ determines the field amplitude and is given in terms of various plasmonic reservoir parameters. $A$ is the quantization area [42].

The length factor $L_{z,\alpha}(\omega)$ is a function of mode frequency $\omega$, and the quantity $AL_{z,\alpha} = V$ is the modes volume that we shall utilize to control emission spectrum and interference. From equations 7 and 8, the length $L_{z,\alpha}$ is given in terms of confinement $\zeta_m = 1/Re[k_m]$ (m=1, 2), and is determined by the physical properties of the NIMM medium namely permittivity's $\varepsilon_{1,2}(\omega)$, and permeability's $\mu_{1,2}(\omega)$, given above. An important point to note here is that large values of the wave number Re [$k_m$] indicate highly confined modes (small $\zeta_m$) at these frequencies and thus highly reduced volume $L_{z,\alpha}$. Likewise suppressed values of Re [$k_m$] lead to poor confinement (large $\zeta_m$) and large value of $L_{z,\alpha}$. It should be also noted here that there is a trade-off between losses and confinement; the more highly confined fields, the more lossy, and the lower the confinement the less lossy the fields. So by appropriate choice of materials, i.e., adjusting the pairs ($\varepsilon_1, \mu_1$) and ($\varepsilon_2(\omega), \mu_2(\omega)$), one can optimize confinement and losses that lead to a decrease in $L_{z,\alpha}$. This reduction in interaction volume provides considerable enhancement of the SP field amplitude $N_\alpha(k_\parallel)$ in Eq. (6). The change in the interaction volume affects the field amplitudes and thus the spontaneous decays and spectrum. This property can be utilized to enhance the coupling between the SP fields and the four-level atomic ensemble as discussed in the next section.

### 4. Atom Interaction with Plasmonic Modes

After quantizing SP fields, we now consider their interaction with four-level atomic system (4LA), as shown in Fig. 1, to explore the quantum interference and its control. Two external classical pumping fields $E_1$ and $E_2$ drive atomic transitions between the states 4 and 1 at transition frequency $\omega_{14}$ and states 4 and 2 at frequency $\omega_{24}$. The upper states 1 and 2 of frequencies $\omega_1$ and $\omega_2$ respectively, spontaneously decay to a common lower state 3 by emission into plasmonic modes of frequency $\omega$ as well as into free space modes. This causes spontaneous emission channels to compete and result in interference effects near NIMM that we like to explore.

The state vector of the system of 4L atom and plasmonic reservoir time evolves according to the Schrodinger equation

$$i\hbar \frac{d}{dt}|\psi(t)\rangle = H_{int}|\psi(t)\rangle \quad . \tag{9}$$

The interaction Hamiltonian $H_{int}$ in the dipole approximation is given by

$$H_{int} = -\boldsymbol{\mu}\cdot\boldsymbol{E} = -\boldsymbol{\mu}_{13}\cdot\hat{\boldsymbol{E}}(r,t) - \boldsymbol{\mu}_{23}\cdot\hat{\boldsymbol{E}}(r,t) - \boldsymbol{\mu}_{14}\cdot\boldsymbol{E}_1(t) - \boldsymbol{\mu}_{24}\cdot\boldsymbol{E}_2(t) \tag{10}$$

where $\boldsymbol{\mu}_{ij}$ is the atomic dipole moment operator connecting levels i and j, i.e.,

$$\boldsymbol{\mu}_{ij} = \langle\boldsymbol{\mu}_{ij}\rangle[\sigma_{ij}(t) + \sigma_{ji}(t)] = \langle\boldsymbol{\mu}_{ij}\rangle[\sigma_{ij}(0)e^{i\omega_{ij}t} + \sigma_{ji}(0)e^{-i\omega_{ij}t}] \quad , \quad \omega_{ij} = \omega_i - \omega_j \quad . \tag{11}$$

Here $\sigma_{ij}(0) = \sigma_{ij} = |i\rangle\langle j|$ are the atomic states projection operators [42], and the driving coherent classical fields are

$$\boldsymbol{E}_l = \boldsymbol{E}_{0l} e^{-i\nu_l t + i\phi_l} + \boldsymbol{E}_{0l}^* e^{i\nu_l t - i\phi_l} \quad , \quad l=1, 2. \tag{12}$$

with frequency $\nu_l$, amplitude $\boldsymbol{E}_{0l}$ and phase $\phi_l$ for the two external fields $l$ =1, 2. The phase difference $\phi_1 - \phi_2 = \pi/4$ will be fixed throughout this paper. The quantized fields $\hat{\boldsymbol{E}}(r,t)$ associated with the plasmonic modes are given in Eqs. (4) – (8) above. Thus, the interaction picture Hamiltonian is given by

$$\begin{aligned}H_{int} &= -\hbar\left(\sum_k g_{1k}\sigma_{13}\hat{a}e^{i\Delta_1 t} + \sum_k g_{1k}^*\sigma_{31}\hat{a}^+ e^{-i\Delta_1 t}\right) - \hbar\left(\sum_k g_{2k}\sigma_{23}\hat{a}e^{i\Delta_2 t} + \sum_k g_{2k}^*\sigma_{32}\hat{a}^+ e^{-i\Delta_2 t}\right) \\ &\quad - \hbar\left(\sigma_{14}\Omega_{14}e^{i\Delta_3 t} + \sigma_{41}\Omega_{14}^* e^{-i\Delta_3 t}\right) - \hbar\left(\sigma_{24}\Omega_{24}e^{i\Delta_4 t} + \sigma_{42}\Omega_{24}^* e^{-i\Delta_4 t}\right)\end{aligned} \tag{13}$$

where

$$\begin{aligned}\Delta_1 &= \omega_{13} - \omega; \quad \Delta_2 = \omega_{23} - \omega = \Delta_1 - \delta_{12} \\ \Delta_3 &= \omega_{14} - \nu_1 \quad ; \Delta_4 = \omega_{24} - \nu_2 \\ \delta_{12} &= \omega_1 - \omega_2.\end{aligned} \tag{14}$$

$$\Omega_{14} = \boldsymbol{\mu}_{14}\cdot\boldsymbol{E}_{01}e^{i\phi_1}/\hbar \quad ; \quad \Omega_{24} = \boldsymbol{\mu}_{24}\cdot\boldsymbol{E}_{02}e^{i\phi_2}/\hbar$$
$$g_{1k} = \boldsymbol{\mu}_{13}\cdot\boldsymbol{E}_k/\hbar \quad ; \quad g_{2k} = \boldsymbol{\mu}_{23}\cdot\boldsymbol{E}_k/\hbar$$

where $\Delta_1$ ($\Delta_2$) is the detuning of the transition frequency $\omega_{13}$ ($\omega_{23}$) from the SP mode frequency $\omega$. Similarly $\Delta_3$ ($\Delta_4$) is the detuning of the transition frequency $\omega_{14}$ ($\omega_{24}$) from the control classical field frequency $\nu_1$ ($\nu_2$), with $\nu_1 = \nu_2$, and $\delta_{12} = \omega_1 - \omega_2$ being the two upper levels frequency difference. $g_{1k}$ and $g_{2k}$ are the coupling strengths of the dipoles $\mu_{13}$ and $\mu_{23}$ to the SP quantized field modes, and $\Omega_{ij}$ is the Rabi frequency coupling the dipole $\mu_{ij}$ to the classical field $E_i$. The state vector $|\psi(t)\rangle$ at time $t$ can be written as

$$|\psi(t)\rangle = a_1(t)|1\rangle|\{0\}\rangle + a_2(t)|2\rangle|\{0\}\rangle + B(t)|4\rangle|\{0\}\rangle + \sum_k C_{3k}(t)|3\rangle|1_k\rangle \tag{15}$$

where $|j\rangle$ (j = 1- 4) is the atomic state, $|\{0\}\rangle$ is the SP field state with no photons, $|1_{k_{\parallel}}\rangle$ is the SP field state with one photon in the mode with polarization $\alpha = TE, TM$ and wave vector $k_{\parallel}$; $|1_k\rangle = a_\alpha^+(k_{//})|0\rangle$, where for notational convenience we set $k=k_{//}$. Here the probability amplitude in level 1 is $a_1$, in level 2 is $a_2$, and in level 3 is $C_{3k}$. Level 4, which has amplitude $B$ is coupled to two upper levels 1 and 2 through the two classical driving fields. The atomic initial states are $a_1(0)$, $a_2(0)$ and $B(0)$.

Using the state vector in Eq. (15) and the interaction Hamiltonian equations (13) into Schrodinger equation (9), we obtain the equations of motion for the probability amplitudes;

$$\dot{a}_1(t) = -\frac{\Gamma_{11}}{2}a_1(t) - \frac{\Gamma_{12}}{2}e^{i\delta_{12}t}a_2(t) + i\Omega_{14}e^{i\Delta_3 t}B(t)$$
$$\dot{a}_2(t) = -\frac{\Gamma_{21}}{2}e^{-i\delta_{12}t}a_1(t) - \frac{\Gamma_{22}}{2}a_2(t) + i\Omega_{24}e^{i\Delta_4 t}B(t) \qquad (16)$$
$$\dot{B}(t) = i\Omega_{14}^* e^{-i\Delta_3 t}a_1(t) + i\Omega_{24}^* e^{-i\Delta_4 t}a_2(t)$$
$$\dot{C}_{3k}(t) = ig_{1k}^* e^{-i\Delta_1 t}a_1(t) + ig_{2k}^* e^{-i\Delta_2 t}a_2(t)$$

Where the quantities $\Gamma_{ij}$ are given [42, 43] by the following expressions,

$$\Gamma_{11} = \sum_k |g_{1k}|^2\, 2\pi\delta(\omega - \omega_{13}) \quad , \quad \Gamma_{22} = \sum_k |g_{2k}|^2\, 2\pi\delta(\omega - \omega_{23})$$
$$\Gamma_{12} = \sum_k g_{1k}g_{2k}^*\, 2\pi\delta(\omega - \omega_{23}) \quad , \quad \Gamma_{21} = \sum_k g_{2k}g_{1k}^*\, 2\pi\delta(\omega - \omega_{13}) \qquad (17)$$

The term $\Gamma_{11}$ gives the spontaneous emission rate into free space and plasmonic modes from level 1 to 3 at the transition frequency $\omega_{13}$ and dipole moment $\mu_{13}$, and likewise $\Gamma_{22}$ gives rate from level 2 to 3 at the transition frequency $\omega_{23}$ and dipole moment $\mu_{23}$. The cross terms $\Gamma_{12}$ and $\Gamma_{21}$ arise from the decay induced coherences between atomic transitions and are responsible for the quantum interference effects [5, 6]. The interference terms depend on the mutual orientations of the diploe moments $\mu_{13}$ and $\mu_{23}$, and are nonzero for non-orthogonal dipoles in vacuum. Therefore, to observe QI in free space the two dipoles have to be near parallel. This requirement can be relaxed by introducing anisotropy [12] in the environment where the dipole is placed. The plasmonic environment we are considering is anisotropic and suits exactly the purpose. The interface that separates the upper and lower half spaces breaks the space isotropy so that radiation rates are different for dipole orientations parallel and normal to interface and this property ensures existence or otherwise of QI effects even for orthogonal dipoles. To implement this scheme [12] we take the atomic levels to be the Zeeman levels with : ( |1>=|j=1, l=-1> , |2>=|j=1, l=1>) and |3>=|j=0, l=0>. The emitted photons from levels |1> and |2> to |3> have left rotating $\boldsymbol{\mu}_{13} = \mu(\hat{z} - i\hat{\mu}_\parallel)$ and right-rotating $\boldsymbol{\mu}_{23} = \mu(\hat{z} + i\hat{\mu}_\parallel)$ polarizations, and so the decay rates and cross terms in equation (17) become;

$$\Gamma_{11} = \Gamma_{22} = \frac{1}{2}(\Gamma_z + \Gamma_\parallel) \quad ; \quad \Gamma_{12} = \Gamma_{21} = \frac{1}{2}(\Gamma_z - \Gamma_\parallel) \quad ; \quad \beta = \frac{\Gamma_z - \Gamma_\parallel}{\Gamma_z + \Gamma_\parallel} \qquad (18)$$

where $\Gamma_z$ is the spontaneous emission rate normal to interface and $\Gamma_\parallel$ is rate parallel to interface and we assumed the two dipoles are equal in magnitudes and transition frequencies $\omega_{13}$ and $\omega_{23}$ are nearly degenerate such that $\delta_{12} = \omega_{13} - \omega_{23} \ll \omega_{13}, \omega_{23}$. The quantity $\beta$ defines the "Degree of QI". The decay rates above in the case of plasmonic structure are calculated, and given by the expressions;

$$\Gamma_z = \Gamma_0 \left(\frac{3\pi}{2}\frac{\mu_z^2}{\mu^2}\right)\frac{c^2}{\omega^2}\frac{c}{|v_{TM}(\omega)|}\frac{k_\parallel^{TM}}{L_{TM}}\left(\left|\frac{k_\parallel^{TM}}{k_1^{TM}}\right|^2\right)e^{-2z/\zeta_1^{TM}} \;;$$

$$\Gamma_\parallel = \Gamma_0 \left(\frac{3\pi}{2}\frac{\mu_\parallel^2}{\mu^2}\right)\frac{c^2}{\omega^2}\left(\frac{c}{|v_{TM}(\omega)|}\frac{k_\parallel^{TM}}{L_{TM}}e^{-2z/\zeta_1^{TM}} + \frac{c}{|v_{TE}(\omega)|}\frac{k_\parallel^{TE}}{L_{TE}}e^{-2z/\zeta_1^{TE}}\right) \qquad (19)$$

The wavenumbers $k_\parallel^{TE,TM}$ for both TE and TM are given by the dispersions equations 1 and 2, the group velocity $v^{TE,TM}$ is calculated from the dispersion relations. The lengths $L_{TE,TM}$ are given from equations 7 and 8, confinement is given by $\zeta = 1/\mathrm{Re}(k_1)$, and $\mu^2 = \mu_\parallel^2 + \mu_z^2$. The decay rate normal to interface receives contribution from TM modes only, while the parallel rate has contributions from both modes. In free space, spontaneous emission rates are isotropic and given by the constant quantity $\Gamma_z = \Gamma_\parallel = \Gamma_o = |\mu_{ij}|^2 \omega_{ij}^3/(3\pi\varepsilon_o\hbar c^3)$ [42], which for optical fields is of order $\Gamma_o = 5\times 10^8 s^{-1}$ and so in free space $\beta = 0$ and no QI can be observed for orthogonal dipoles. In cavity or structured environment, the spontaneous emission rate is anisotropic and can be modified by the environment in which the atom is embedded as was first observed by the pioneering work of Purcell [43], which paved the way for the full branch of research of cavity quantum electrodynamics [44-45]. Thus for orthogonal diploes the environment anisotropy ensures the existence of QI effects. Furthermore, as we see from Eqns. (16), the interference cross terms are accompanied by the frequency difference of the two upper levels $\delta_{12} = \omega_1 - \omega_2 \ll \omega_{13}, \omega_{23}$. The emission rates, dipole orientations, upper levels frequency difference, and environment, all affect the pattern of interference and spectrum, and provide additional control near the NIMM interface.

The decay rates are given in terms of free space rate and the quantities $\Gamma_{z,\parallel} / \Gamma_0$ define the Purcell enhancement factors of the decay rates in structured environment relative to free space case. In Fig.3 we show Re[$k_1$] and the coupling strength $|g_{1k}(z,\omega)| = (\mu/\hbar\sqrt{2})\sqrt{|E_z(z,\omega)|^2 + |E_\parallel(z,\omega)|^2}$ between the SP fields and atom as functions of mode frequency, whereas spontaneous emission rate Purcell factors are shown in Fig.4 as $\Gamma_{z,\parallel} / \Gamma_0$, along with $\beta$ (degree of QI). The enhancement of the coupling strength and spontaneous emission rate, can be traced to the basic SP properties of the NIMM structure and can be made more clearly through the joint effect of basic parameters; the wavenumber $k_\parallel$ that enters in the dispersion relation, the real parts of $k_1$ and $k_2$ that give confinement factor $L_{TM,TE}$ in Eqs. (7) and (8), which affects the interaction volume, and the plasmonic modes group velocity factors $\upsilon_{TE,TM} = \partial \omega(k_\parallel) / \partial k_\parallel$. As seen in Figs.2b, the dispersion curves flatten as the wavenumber gets larger and foldings occur near the mode frequencies $\omega / \omega_e = 0.4, 0.6, 0.65$. The values of $k_\parallel$ are very large at 0.4 and 0.65, while at 0.6 it is very small and approaches zero. The corresponding values of the SP group velocity at these mode frequencies are highly reduced and SP are slowed down with much slower TE modes. Furthermore, the factor $L_{TM,TE}$ is inversely proportional to real $k_1$ and $k_2$ according to Eqs. (7) and (8) and its behavior is determined from that of $k_1$ and $k_2$. Both the decay rates and coupling strength strongly depend on confinement ($\zeta_1 = 1/\text{Re}[k_1]$) and on the position of the mode frequency with respect to the dispersion curves. For example in Fig.3a, at mode frequency $\omega / \omega_e \approx 0.42$, both TE and TM modes show very small values of real $k_1$ and thus poor confinement (large $\zeta_1 = 1/\text{Re}[k_1]$) and large $L_{TE,TM}$, and thus highly suppressed decay rates. At frequencies slightly smaller or bigger than 0.42, TE (TM) real $k_1$ is very large (small) which leads to good (poor) confinement and thus reduced (large) $L_{TE,TM}$. This translates into enhancement (suppression) of parallel (normal) decay rates.

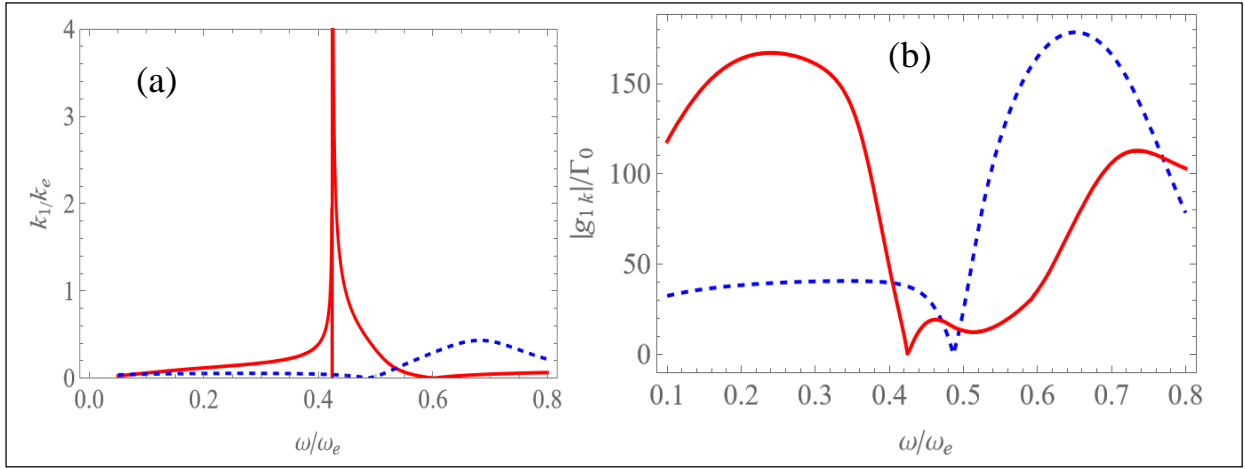

Fig.3. (a) Real ($k_1$) in units of $k_e$ in medium 1 for TE mode (solid) and TM mode (dashed), (b) the atom-SP coupling strength (in units of $\Gamma_o$) as functions of SP frequency, atom is placed 10nm above interface. Dipole is normal (dashed) and parallel (solid) to interface.

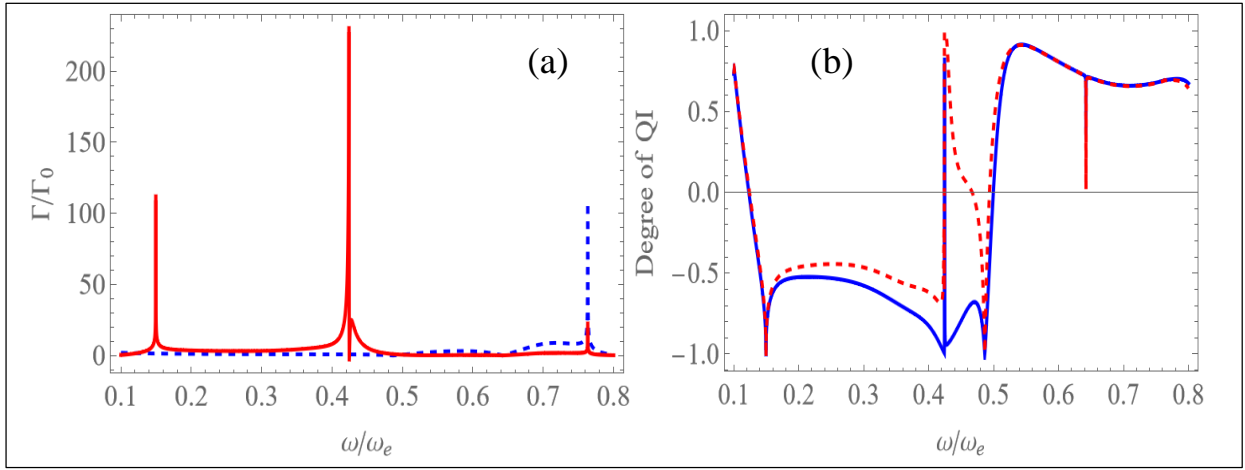

Fig.4. (a) The spontaneous emission rates (in units of $\Gamma_o$) into SP modes for dipole parallel to interface (solid) and normal to interface (dashed) for atom placed at z=10nm above interface. (b) Degree of QI for atom placed at z=10nm (solid) and z=40nm (dashed) lines.

The decay rates show peaks at frequencies $\omega/\omega_e = 0.15$ and $0.76$, which can be explained in the same way due to behavior of confinement. Therefore, the net product of these factors $k_\parallel$, $L_{TM,TE}$, and $\left|\partial\omega(k_\parallel)/\partial k_\parallel\right|$, explains the behavior of the coupling and decay rates and its enhancement few orders of magnitudes at moderate set of parameters, compared to free space case.

Furthermore, the slow propagation of SP modes can be regarded as another temporal "longitudinal confinement" along the propagation direction in addition to field spatial confinement normal to propagation direction. The total effect of this confinement is to reduce the interaction volume and increase the interaction time with the atom. This results in the enhanced decays.

Fig.4b shows the degree of QI ($\beta$) which is directly related to the decay rates we just discussed. The normal decay rates receive contribution from TM modes only while both TM and TE modes contribute to the parallel rates. In frequency regions where normal decay is larger than the parallel rates $\beta$ is positive, whereas for parallel rates larger than normal rates $\beta$ is negative. For frequency $\omega/\omega_e \approx 0.42$, TE decay rate is very high and values of $\beta$ are negative. This frequency region however falls within the high loss region to which we alluded earlier. The negative values of $\beta$ are attributed to TE modes contribution being higher. Similar results for negative $\beta$ due to TE have also been discussed in the literature [17] using similar artificial materials.

## 5. Spontaneous spectrum for SP modes

Having discussed QI and its measure, we now explore these effects in the NIMM plasmonic structure numerically using spectrum function. First, we solve the system of Eq. (16) to get

$$a_1(t) = \sum_j \alpha_j e^{\lambda_j t} \quad ; \quad a_2(t) = e^{-i\delta_{12}t}\sum_j f_j \alpha_j e^{\lambda_j t} \quad , \quad B(t) = e^{-i\Delta_3 t}\sum_j h_j \alpha_j e^{\lambda_j t} \quad (j=1,2,3), \quad (20)$$

and the probability amplitude $C_{3k}(t)$ is readily obtained from Eq. (20), where $\lambda_j$ are the complex roots of the cubic equation $\lambda^3 + a\lambda^2 + b\lambda + c = 0$, and the coefficients $\alpha_j, f_j, h_j$ and $a,b,c$ are given in Appendix B.

The spectrum function describes the frequency distribution of radiation and is defined as the Fourier transform of the time correlations of the emitted fields and given by the expression [10-11, 42];

$$S(\omega_{ij}) = \Gamma_{ij}/(2\pi|g_{ik}|^2)|C_{3k}(t\to\infty)|^2 \qquad (21)$$

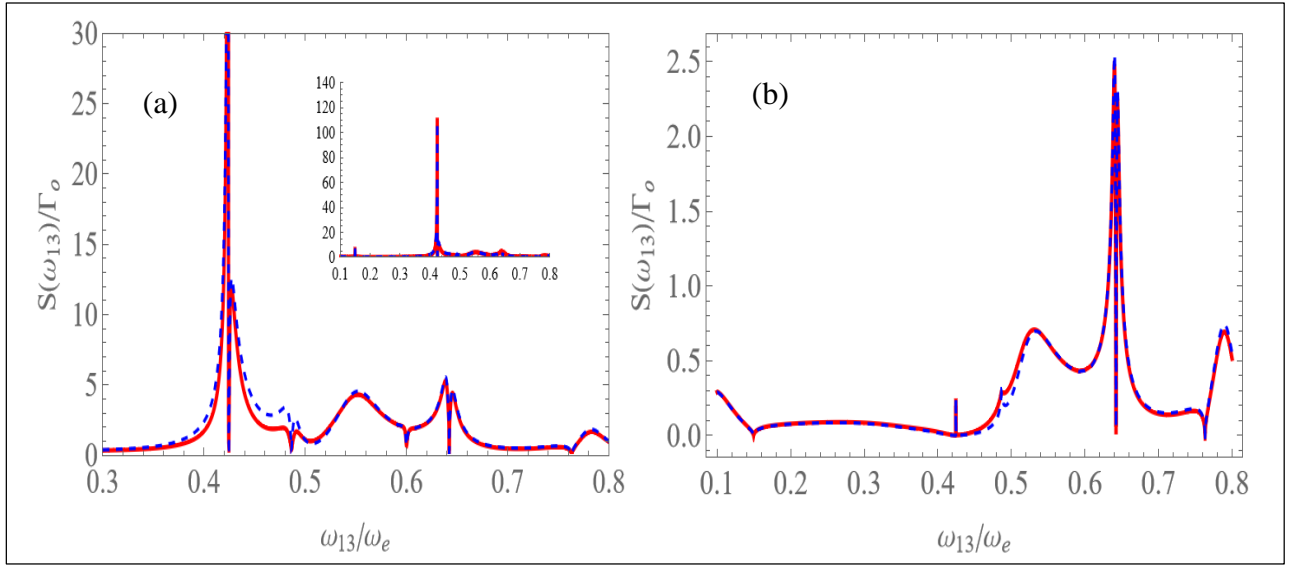

Fig.5. Spontaneous emission spectrum in units of $\Gamma_o$ for orthogonal dipoles as a function of scaled atomic transition frequency $\omega_{13}/\omega_e$ for atom locations $z=10nm$ (solid) and $z=40nm$ (dashed) above interface and mode frequency $\omega = 0.6\omega_e$. The parameters are; $a_1(0) = 1, a_2(0) = 0, B_0 = 0, \Delta_3 = 0, \delta_{12} = 0.001\Gamma_o$, (a) Rabi frequency $|\Omega_{14}| = |\Omega_{24}| = 0.02\Gamma_o$, (b) $|\Omega_{14}| = |\Omega_{24}| = 4.5\Gamma_o$. Inset (a) shows full scale graph.

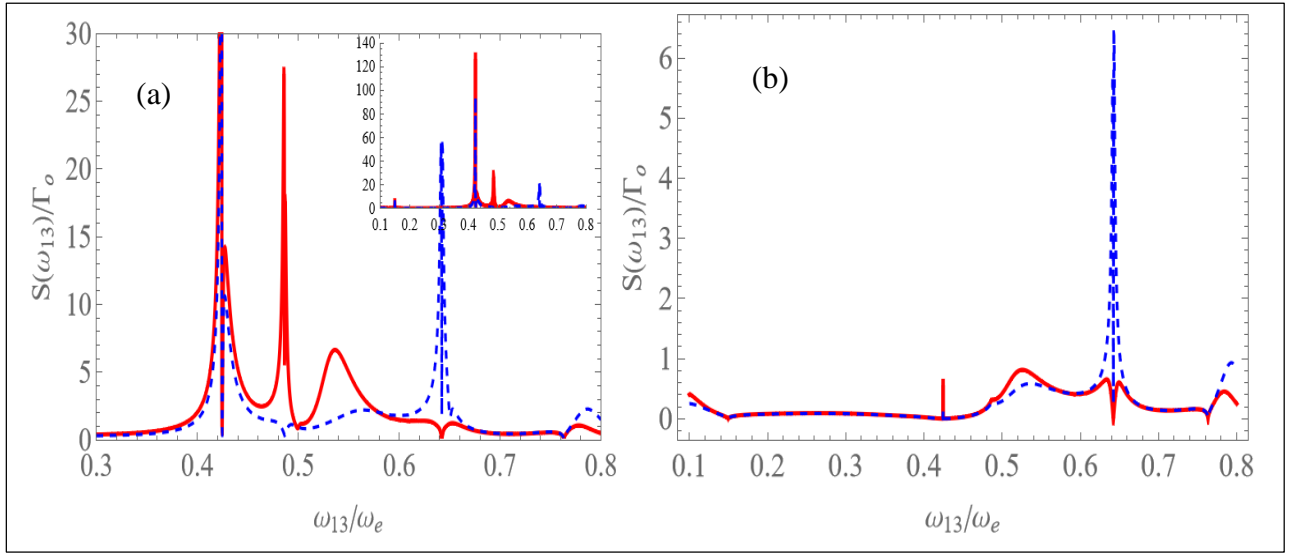

Fig.6. Spontaneous emission spectrum for orthogonal dipoles as a function of scaled atomic transition frequency $\omega_{13}/\omega_e$ for mode frequency $\omega = 0.5$ (solid) and $\omega = 0.65\omega_e$ (dashed). Atom is at z=10nm. The parameters are as in Fig.5 a, b. Inset shows full scale graph.

This spectrum function is shown in Figs.5 and 6 for an atom placed near the NIMM interface as a function of transition frequency $\omega_{13}/\omega_e$ scanning the plasmonic mode frequency for varied atomic locations, varied mode frequency and for varied Rabi frequency.

The emission spectrum is very much dependent on the emission rates, among many other parameters, and its behavior is determined from the relative size of the different parameters.

To see this in more details we note that the spectrum function at the transition frequency $\omega_{ij}$ and the decay rate $\Gamma_{ij}$, takes the general form $S(\omega_{ij}) \propto \Gamma_{ij}/[\Gamma_{ij}^2 + R(\Delta, \Omega, ...)]$ where $R(\Delta, \Omega, ...)$ is a function of various quantities like detuning $\Delta$, Rabi frequency $\Omega$, and SP reservoir parameters. For strong Rabi frequency, ($\Omega > \Gamma_{ij}$), and off resonance, the function R is appreciable and the spectrum behaves like $S(\omega_{ij}) \propto \Gamma_{ij}$ (Fig.5a and Fig.6a). Near resonances and in the weak Rabi frequency case ($\Omega \ll \Gamma_{ij}$), the function R is usually small and the spectrum in this case behaves like $S(\omega_{ij}) \propto \Gamma_{ij}^{-1}$ (Fig.5b and Fig.6b). So in the strong Rabi case the spectrum is directly related to the behavior of emission rates (Fig.4a) which shows peaks or dips matching those of the decay rates. The frequency $\omega/\omega_e = 0.6$ in Fig.5a is the sample field mode and resonance with atom occurs at $\omega_{13} \approx \omega$. At this frequency $\omega/\omega_e = 0.6$, interference effects lead to destructive interference and constructive interference along the two side of this frequency leading to dark line. In Fig.6a, atom-field resonance occurs at frequencies $\omega/\omega_e = 0.5, 0.65$. Figs.5, 6b refer to the case where the Rabi frequency is weak and the spectrum behaves like $S(\omega_{ij}) \propto \Gamma_{ij}^{-1}$, which can be explained in the same way. The atom-field resonance peaks slightly shift position in a manner that depends on the strength of Rabi frequency, detuning and location with respect to dispersion curves. We note that the behavior of quantum interference effects depend strongly on the transition and mode frequency relative to the dispersion relations. We should also remark that these QI effects occur in the frequency regions in accordance with the QI measure $\beta > 0$.

## 5. Conclusion and discussion

We proposed an anisotropic plasmonic NIMM environment that supports both TE and TM polarized surface plasmon modes to enhance their coupling to a four-level atom spontaneous

arXiv:2309.10710V2 [physics.optics] 19 Sept 2023emission rates and spectrum. The anisotropy of this environment facilitates QI for orthogonal diploes with rich parameters. This rich set of media and atomic parameters provides more flexibility and options for controlling emission spectrum. The emission rates and spectrum have been investigated for various atomic and media parameters, such as the location of atom, transition frequency relative to the dispersion curves, Rabi frequency etc. This shows considerable effects on the interference terms and spectrum leading to two orders of magnitudes enhancement. Such enhancements are comparable with other works in recent literature [17, 19, 24, 46], and surpass it in some cases, thus supporting this NIMM plasmonic environment proposal to enhance spontaneous emission and spectrum and other atom-field coupling processes.

We demonstrated that plasmonic modes engineered near NIMM interface generate strong coupling to emitters placed near interface and can induce rich interference effects, and thus SP modes with prescribed properties are viable candidates for many nano-photonics and plasmonics applications. With current technology, NIMMs have reached optical frequencies [47] and NIMMs at 780nm wavelength [48] and visible wavelength of 580 nm [49] have been attained. These NIMMs at the aforthmentioned optical wavelengths are commensurate, respectively, with atomic transition wavelengths of 780nm for Rb87 atoms and Sodium 589nm. Therefore, implementation of this scheme with atoms at NIMM interface at right optical wavelength should be possible.

**Appendix A**: Derivation of normalization factors of Eqs. (7) and (8).

We write the electromagnetic field Hamiltonian in a dispersive medium [50, 51] as

$$H_{field} = \frac{1}{2}\int d^3 r \left[ \tilde{\varepsilon}|\hat{E}(r)|^2 + \tilde{\mu}|\hat{H}(r)|^2 \right]$$
$$\tilde{\varepsilon} = \text{Re}\left(\frac{\partial}{\partial \omega}[\omega \varepsilon_o \varepsilon(\omega)]\right) \quad , \quad \tilde{\mu} = \text{Re}\left(\frac{\partial}{\partial \omega}[\omega \mu_o \mu(\omega)]\right)$$

(A1)

The SP electric field operator is constructed as sum of SP modes, and making the usual prescription $\sum_{\alpha,k_{\parallel}} \to A/(2\pi)^2 \sum_\alpha \int d^2 k_{\parallel}$, the SP field is written as;

$$\hat{E}(r) = \frac{A}{4\pi^2}\sum_\alpha \int d^2 k_{\parallel} [E_\alpha(k_{\parallel}) a_\alpha(k_{\parallel}) e^{i(k_{\parallel} \cdot r_{\parallel} - \omega t)} + H.C]$$

(A2)

where $A$ is the quantization area. The plasmonic modes annihilation and creation operators, $a_\alpha(k_\parallel)$ and $a_\alpha^+(k_\parallel)$ of the mode $\alpha = TE, TM$ and wave vector $k_\parallel$ obey the usual equal time commutation relation $[a_\alpha(k_\parallel), a_{\alpha'}^+(k_\parallel')] = \delta_{\alpha\alpha'}\delta(k_\parallel - k_\parallel')$. The mode functions satisfying the wave equation are;

$$\boldsymbol{E}_{TE}(\boldsymbol{k}_\parallel) = N_{TE}(\boldsymbol{k}_\parallel)(\hat{\boldsymbol{k}}_\parallel \times \hat{\boldsymbol{z}})\left[\theta(z)e^{-k_1 z} + \theta(-z)e^{k_2 z}\right]$$

$$\boldsymbol{E}_{TM}(\boldsymbol{k}_\parallel) = N_{TM}(\boldsymbol{k}_\parallel)\left[\theta(z)(\hat{\boldsymbol{k}}_\parallel + i\hat{\boldsymbol{z}}\frac{k_\parallel}{k_1})e^{-k_1 z} + \theta(-z)(\hat{\boldsymbol{k}}_\parallel - i\hat{\boldsymbol{z}}\frac{k_\parallel}{k_2})e^{k_2 z}\right]$$
(A3)

The corresponding magnetic field operator $\hat{H}$ is determined from Maxwell equation $\nabla \times \hat{E} = i\omega\mu_o\mu(\omega)\hat{H}$. The SP field amplitudes $N_\alpha$ ($\alpha$=TM and TE), are determined by the requirement that the field Hamiltonian in dispersive medium (Eq.A1) reduces to the canonical form Hamiltonian

$$H_{field} = \frac{1}{2}\frac{A}{4\pi^2}\sum_\alpha \int d^2 \boldsymbol{k}_\parallel \hbar\omega(k_\parallel)[\hat{a}_\alpha(k_\parallel)\hat{a}_\alpha^+(k_\parallel) + \hat{a}_\alpha^+(k_\parallel)\hat{a}_\alpha(k_\parallel)]$$
(A4)

Now we use the fields in Eqs A2-A3 into the Hamiltonian A1 to evaluate the space integrals in A1. Details are given for the TE case only. From Eq A1;

$$H_{field} = \frac{1}{2}\int d^3r\left[\tilde{\varepsilon}|\hat{E}(r)|^2 + \tilde{\mu}|\hat{H}(r)|^2\right]$$

$$= \frac{1}{2}\int d^2r_\parallel \left\{\left[\int_0^\infty dz\tilde{\varepsilon}_1|\hat{E}(z>0)|^2 + \int_{-\infty}^0 dz\tilde{\varepsilon}_2|\hat{E}(z<0)|^2\right] + \left[\int_0^\infty dz\tilde{\mu}_1|\hat{H}(z>0)|^2 + \int_{-\infty}^0 dz\tilde{\mu}_2|\hat{H}(z<0)|^2\right]\right\}$$
(A5)

Keeping only the energy conserving terms we have for the E-part of integrals;

$$I_{E1} = \int d^2 r_{\parallel} \int_0^{\infty} dz \tilde{\varepsilon}_1 |\hat{E}(z>0)|^2$$

$$= |N_{TE}|^2 \tilde{\varepsilon}_1 \int d^2 r_{\parallel} \int_0^{\infty} dz\, e^{(-2\text{Re}[k_1]z)} \left[ \sum_{\alpha,\alpha'} \int d^2 k_{\parallel} d^2 k_{\parallel}' \hat{a}_{\alpha}(k_{\parallel}) \hat{a}_{\alpha'}^+(k_{\parallel}') e^{i[(k_{\parallel}-k_{\parallel}')\cdot r_{\parallel} - (\omega-\omega')t]} + H.C \right]$$

$$= |N_{TE}|^2 \frac{(2\pi)^2 \tilde{\varepsilon}_1}{2\text{Re}[k_1]} \sum_{\alpha} \int d^2 k_{\parallel} \left[ \hat{a}_{\alpha}(k_{\parallel}) \hat{a}_{\alpha}^+(k_{\parallel}) + \hat{a}_{\alpha}^+(k_{\parallel}) \hat{a}_{\alpha}(k_{\parallel}) \right]$$

$$I_{E2} = \int d^2 r_{\parallel} \int_{-\infty}^{0} dz \tilde{\varepsilon}_2 |\hat{E}(z<0)|^2$$

$$= |N_{TE}|^2 \tilde{\varepsilon}_2 \int d^2 r_{\parallel} \int_{-\infty}^{0} dz\, e^{(2\text{Re}[k_2]z)} \left[ \sum_{\alpha,\alpha'} \int d^2 k_{\parallel} d^2 k_{\parallel}' \hat{a}_{\alpha}(k_{\parallel}) \hat{a}_{\alpha'}^+(k_{\parallel}') e^{i[(k_{\parallel}-k_{\parallel}')\cdot r_{\parallel} - (\omega-\omega')t]} + H.C \right]$$

$$= |N_{TE}|^2 \frac{(2\pi)^2 \tilde{\varepsilon}_2}{2\text{Re}[k_2]} \sum_{\alpha} \int d^2 k_{\parallel} \left[ \hat{a}_{\alpha}(k_{\parallel}) \hat{a}_{\alpha}^+(k_{\parallel}) + \hat{a}_{\alpha}^+(k_{\parallel}) \hat{a}_{\alpha}(k_{\parallel}) \right]$$

(A6)

Adding the electric part of energy, we have

$$\int d^3 r \tilde{\varepsilon} |\hat{E}(r)|^2 = \varepsilon_o (2\pi)^2 |N_{TE}|^2 D_{TE} \sum_{\alpha} \int d^2 k_{\parallel} \left[ \hat{a}_{\alpha}(k_{\parallel}) \hat{a}_{\alpha}^+(k_{\parallel}) + \hat{a}_{\alpha}^+(k_{\parallel}) \hat{a}_{\alpha}(k_{\parallel}) \right]$$

$$D_{TE} = \left[ \frac{\text{Re}(\partial_{\omega}[\omega \varepsilon_1(\omega)])}{2\text{Re}[k_1]} + \frac{\text{Re}(\partial_{\omega}[\omega \varepsilon_2(\omega)])}{2\text{Re}[k_2]} \right]$$

(A7)

where $\partial_{\omega} = \partial/\partial \omega$. Similarly the magnetic part of energy for the TE case is

$$I_{H1} = \int d^2 r_{\parallel} \int_0^{\infty} dz\, \tilde{\mu}_1 |H(z>0)|^2$$

$$= |N_{TE}|^2 \tilde{\mu}_1 \frac{|k_1|^2 + |k_{\parallel}|^2}{\omega^2 \mu_o^2 |\mu_1^2|} \int d^2 r_{\parallel} \int_0^{\infty} dz\, e^{(-2\text{Re}[k_1]z)} \left[ \sum_{\alpha,\alpha'} \int d^2 k_{\parallel} d^2 k_{\parallel}' \hat{a}_{\alpha}(k_{\parallel}) \hat{a}_{\alpha'}^+(k_{\parallel}') e^{i[(k_{\parallel}-k_{\parallel}')\cdot r_{\parallel} - (\omega-\omega')t]} + H.C \right]$$

$$= \varepsilon_o (2\pi)^2 |N_{TE}|^2 \frac{c^2}{\omega^2} \frac{|k_1|^2 + |k_{\parallel}|^2}{|\mu_1^2|} \frac{\text{Re}(\partial_{\omega}[\omega \mu_1(\omega)])}{2\text{Re}[k_1]} \sum_{\alpha} \int d^2 k_{\parallel} \left[ \hat{a}_{\alpha}(k_{\parallel}) \hat{a}_{\alpha}^+(k_{\parallel}) + \hat{a}_{\alpha}^+(k_{\parallel}) \hat{a}_{\alpha}(k_{\parallel}) \right]$$

(A8)

$$I_{H2} = \int d^2 r_{\parallel} \int_{-\infty}^{0} dz\, \tilde{\mu}_2 |H(z<0)|^2$$

$$= |N_{TE}|^2 \tilde{\mu}_2 \frac{|k_2|^2 + |k_{\parallel}|^2}{\omega^2 \mu_o^2 |\mu_2^2|} \int d^2 r_{\parallel} \int_{-\infty}^{0} dz\, e^{(2\text{Re}[k_2]z)} \left[ \sum_{\alpha,\alpha'} \int d^2 k_{\parallel} d^2 k_{\parallel}' \hat{a}_{\alpha}(k_{\parallel}) \hat{a}_{\alpha'}^+(k_{\parallel}') e^{i[(k_{\parallel}-k_{\parallel}')\cdot r_{\parallel} - (\omega-\omega')t]} + H.C \right]$$

$$= \varepsilon_o (2\pi)^2 |N_{TE}|^2 \frac{c^2}{\omega^2} \frac{|k_2|^2 + |k_{\parallel}|^2}{|\mu_2^2|} \frac{\text{Re}(\partial_{\omega}[\omega \mu_2(\omega)])}{2\text{Re}[k_2]} \sum_{\alpha} \int d^2 k_{\parallel} \left[ \hat{a}_{\alpha}(k_{\parallel}) \hat{a}_{\alpha}^+(k_{\parallel}) + \hat{a}_{\alpha}^+(k_{\parallel}) \hat{a}_{\alpha}(k_{\parallel}) \right]$$

Adding the magnetic part of energy, we have

$$\int d^3r \, \tilde{\mu} |\hat{H}(r)|^2 = \varepsilon_o (2\pi)^2 |N_{TE}|^2 \frac{c^2}{\omega^2} S_{TE} \sum_\alpha \int d^2k_\parallel [\hat{a}_\alpha(k_\parallel)\hat{a}_\alpha^+(k_\parallel) + \hat{a}_\alpha^+(k_\parallel)\hat{a}_\alpha(k_\parallel)]$$

$$S_{TE} = \left[ \frac{|k_1|^2 + |k_\parallel|^2}{|\mu_1^2|} \frac{\text{Re}(\partial_\omega[\omega\mu_1(\omega)])}{2\text{Re}[k_1]} + \frac{|k_2|^2 + |k_\parallel|^2}{|\mu_2^2|} \frac{\text{Re}(\partial_\omega[\omega\mu_2(\omega)])}{2\text{Re}[k_2]} \right]$$

(A9)

So the total Hamiltonian in the plasmonic environment in Eq. (A5) becomes

$$H_{field} = \frac{1}{2}\varepsilon_o (2\pi)^2 |N_{TE}|^2 L_{TE} \left(\frac{A}{4\pi^2}\right)^2 \sum_\alpha \int d^2k_\parallel [\hat{a}_\alpha(k_\parallel)\hat{a}_\alpha^+(k_\parallel) + \hat{a}_\alpha^+(k_\parallel)\hat{a}_\alpha(k_\parallel)]$$

$$L_{TE} = \left[ D_{TE} + \frac{c^2}{\omega^2} S_{TE} \right].$$

(A10)

This field Hamiltonian reduces to the canonical Hamiltonian A4, when

$$\varepsilon_o A L_{TE} |N_{TE}|^2 = \hbar\omega(k_\parallel) \ . \tag{A11}$$

which is Eq. 6. Similar calculations for the TM case lead to Eqs. 6 and 8.

**Appendix B: Coefficients in Eq. (20).**

In Eq. (18), $\lambda_j$ are the roots of the cubic equation $\lambda^3 + a\lambda^2 + b\lambda + c = 0$, with

$$a = \frac{1}{2}(\Gamma_{11} + \Gamma_{22}) - i(\Delta_3 + \delta_{12})$$

$$b = \frac{1}{4}(\Gamma_{11}\Gamma_{22} - \Gamma_{12}\Gamma_{21}) + |\Omega_{14}|^2 + |\Omega_{24}|^2 - i\Delta_3 \frac{1}{2}(\Gamma_{11} + \Gamma_{22}) - \delta_{12}(\Delta_3 + \frac{i}{2}\Gamma_{11})$$

$$c = \frac{1}{2}(\Gamma_{11}|\Omega_{24}|^2 + \Gamma_{22}|\Omega_{14}|^2) - \frac{1}{2}(\Gamma_{12}\Omega_{24}\Omega_{14}^* + \Gamma_{21}\Omega_{14}\Omega_{24}^*)$$

$$-\frac{i}{4}\Delta_3(\Gamma_{11}\Gamma_{22} - \Gamma_{12}\Gamma_{21}) - \delta_{12}(\frac{1}{2}\Gamma_{11}\Delta_3 + i|\Omega_{14}|^2)$$

(B1)

and the probability amplitude coefficients $\alpha_j (j=1,2,3)$ are given as

$$\alpha_1 = \frac{B_o(f_3 - f_2) + a_{20}(h_2 - h_3) + a_{10}(f_2 h_3 - f_3 h_2)}{f_1(h_2 - h_3) + f_2(h_3 - h_1) + f_3(h_1 - h_2)},$$

$$\alpha_2 = \frac{B_o(f_1 - f_3) + a_{20}(h_3 - h_1) + a_{10}(f_3 h_1 - f_1 h_3)}{f_1(h_2 - h_3) + f_2(h_3 - h_1) + f_3(h_1 - h_2)}, \quad (B2)$$

$$\alpha_3 = \frac{B_o(f_2 - f_1) + a_{20}(h_1 - h_2) + a_{10}(f_1 h_2 - f_2 h_1)}{f_1(h_2 - h_3) + f_2(h_3 - h_1) + f_3(h_1 - h_2)}.$$

$$f_j = \frac{\lambda_j + \mu}{\nu + (\lambda_j - i\delta_{12})\left(\frac{\Omega_{14}}{\Omega_{24}}\right)} \quad , \quad h_j = \frac{\Omega_{14}^* + \Omega_{24}^* f_j}{-i\lambda_j - \Delta_3} \quad (B3)$$

$$\mu = \frac{1}{2}\left(\Gamma_{11} - \Gamma_{21}\frac{\Omega_{14}}{\Omega_{24}}\right) \quad , \quad \nu = \frac{1}{2}\left(-\Gamma_{12} + \Gamma_{22}\frac{\Omega_{14}}{\Omega_{24}}\right). \quad (B4)$$